# Simulation of Nonstationary Electromagnetic Scattering Using the Time- and Frequency-Domain Approaches

I. G. Efimova

Moscow State University

*Abstract.* Scattering of nonstationary electromagnetic fields from axially symmetrical bodies is numerically investigated. Simulations are performed using the time- and frequency-domain approaches. Computational results obtained for a finite perfectly conducting cylinder, perfectly conducting cone, and cone covered with a dielectric layer are analyzed.

INTRODUCTION

Scattering of nonstationary electromagnetic fields can be studied using either the time- or frequency domain approach. The time-domain methods are more suitable in the case when different bodies are successively illuminated by one and the same pulse and it is necessary to investigate, for example, the effect of the shape of a scatterer on the response. The time-domain approach applied in this paper involves solution of the integro-differential equations for the current flowing over the surface of a scatterer.

The frequency-domain methods are more efficient in the analysis of scattered fields when various incident pulses are scattered by the same body. Then, one can calculate the frequency response of a body only once and apply the Fourier transform to the product of the response and the spectrum of each incident signal, which appears to be a good tool to study the effect of the incident pulse on the scattered field. Below, application of both approaches is exemplified by simulations of electromagnetic scattering of different pulses from various axially symmetrical bodies.

1. SCATTERING FROM A PERFECTLY CONDUCTING FINITE CYLINDER

Let us consider a perfectly conducting circular cylinder of a finite length. The cylinder is illuminated by a Gaussian video pulse propagating perpendicularly to the symmetry axis of the body. In this case, the induced current has the maximum surface density. When the electric (magnetic) field is aligned with the axis, the transverse (longitudinal) current is absent on the lateral surface of the cylinder. Computations were performed using the numerical algorithm developed for solving the problem of transient scattering from bodies of revolution [1, 2]. Figure 1 shows the current calculated at the center point of the generatrix on the illuminated part of the surface for both polarizations of the incident field. The following notation is introduced: $J_t$ and $J_l$ are the transverse and longitudinal current components, respectively; $L$ and $a$ are the length and radius of the cylinder, respectively; $\tau$ is the pulse duration; $t$ is time; and $c$ is the velocity of light in free space.

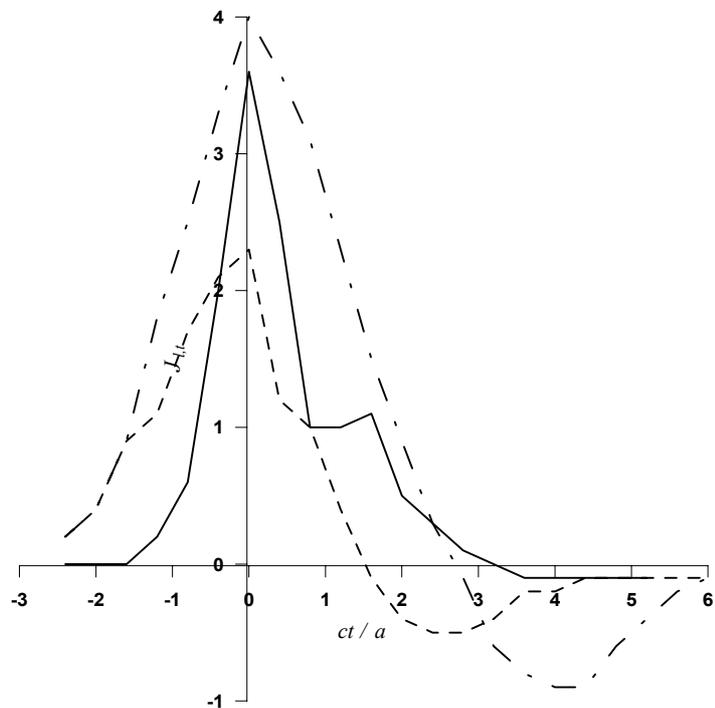

Fig. 1

Figure 1 shows the longitudinal current components calculated at $c\tau / a = 1$ and $\tau = 3.35$ ns (solid line) and at $c\tau / a = 2.25$ and $\tau = 7.5$ ns (dotted-and-dashed line) as well as the transverse component calculated $c\tau / a = 2.25$ and $\tau = 7.5$ ns (dashed line); here, $L / a = 5$. The first two curves are obtained in the case when the incident electric field is aligned with the axis of symmetry, and the third curve is obtained for the incident electric field perpendicular to the axis.

2. SCATTERING FROM A PERFECTLY CONDUCTING CONE

Let a finite perfectly conducting circular cone be illuminated by a rectangular radio or video pulse propagating along the axis of symmetry towards the vertex. To solve numerically the scattering problem, we apply the frequency-domain approach involving the method of frequency-domain integral equations for axially symmetric bodies and the fast Fourier transform algorithm [3]. Since the calculation technique is well known, we focus here on computational results and analyze the physical effects observed.

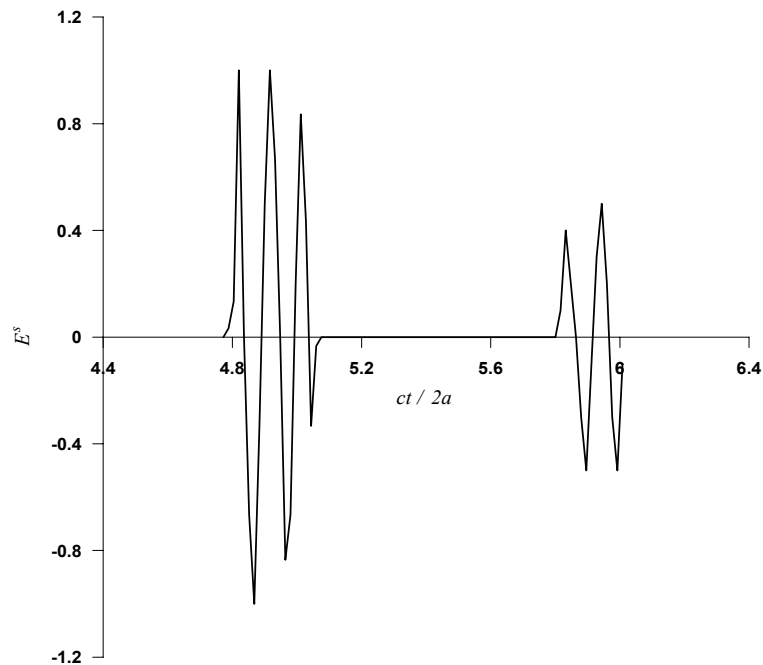

Fig. 2.

Figures 2 and 3 show the time dependences of the far-zone scattered fields in the cases when $c\tau / 2\alpha = 0.25$ and $\omega\tau / 2\pi = 2$ (Fig. 2) and 0 (Fig. 3), where $\omega$ is the carrier frequency (the angular opening of the cone is 23°). The scattered signal contains two pulses corresponding to the first- and second-order diffraction from the edge of the cone base. It is seen that the shape of the radio pulse envelope is restored virtually in one or two periods of the carrier-frequency oscillations.

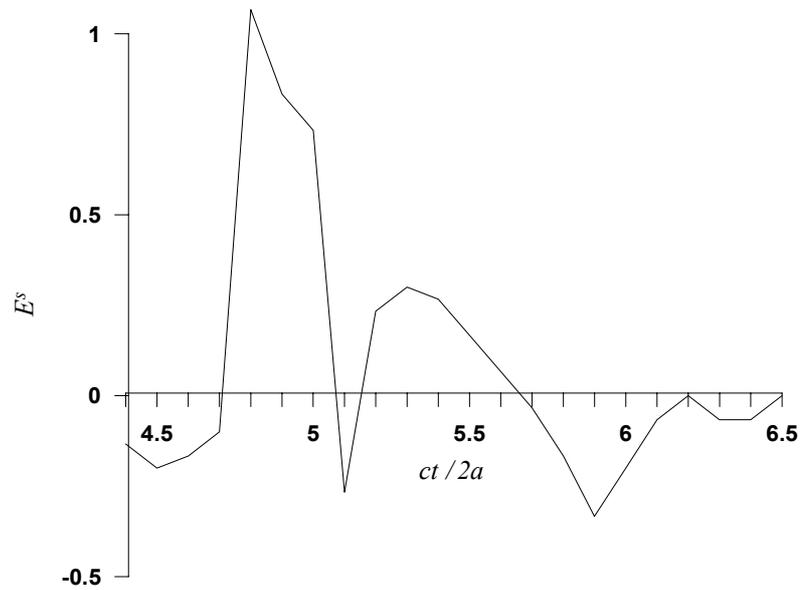

Fig. 3.

When the carrier frequency is equal to zero, a video pulse is scattered. At $c\tau/2a = 0.25$, the scattered signal has a complex shape. The pulse with the larger amplitude is due to the first-order diffraction from the cone base edge. Its duration is virtually equal to that of the incident pulse. The pulse due to the second-order diffraction has the smaller amplitude. One can see that the shape of the reflected video signal fundamentally differs from that of the envelope of the reflected rectangular incident pulse.

3. SCATTERING FROM A PERFECTLY CONDUCTING CONE WITH A DIELECTRIC COATING

We consider a perfectly conducting cone covered with a dielectric layer of a fixed thickness. The dielectric, characterized by permittivity $\varepsilon$, is assumed to be homogeneous. As in Section 2, we apply the frequency-domain approach involving the method of integral equations and the fast Fourier transform algorithm [4]. The incident field is assumed to be a Gaussian video pulse propagating along the cone axis towards the vertex. Figure 4 shows the time dependences of the scattered field obtained at ε = 1 (solid curve), 2 (dashed curve), and 4 (dashed-and-dotted curve).

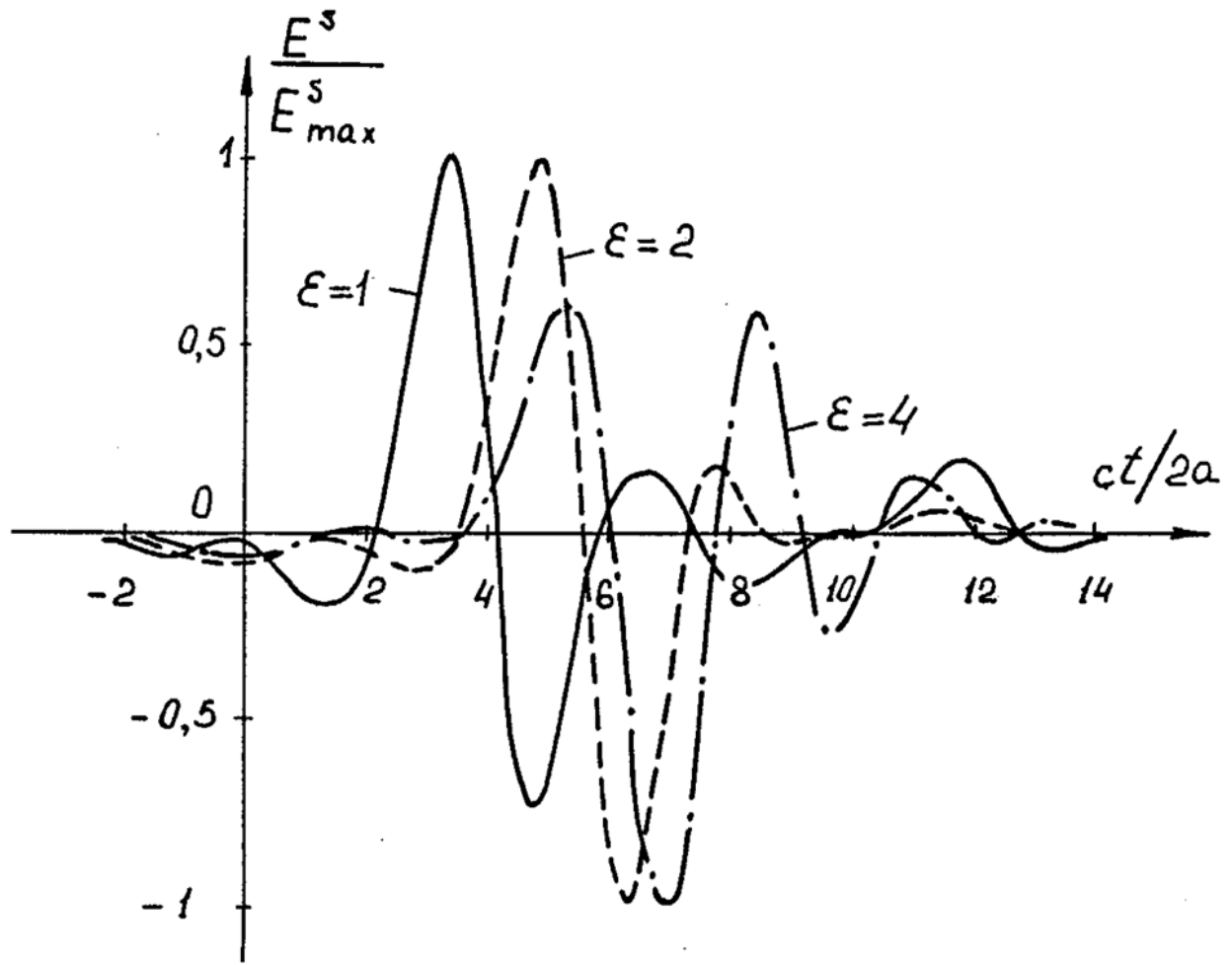

It is seen that, similarly, to the scattering of a rectangular video or radio pulse by a perfectly conducting cone, one can also separate the fields due to the first- and second-order diffraction from the edge of the metal base. However, when the incident wave travels towards the vertex, the presence of the dielectric layer causes certain changes in the backscattered field. First of all, the amplitude of the field primarily diffracted from the edge of the metal base increases. The reason is that, in the presence of a dielectric layer, a considerable portion of the energy of this field is transformed into the energy of a quasi-surface wave, which propagates in the layer on the base and is again scattered by the edge. As the permittivity increases, this portion of energy increases initiating the pulses of higher order diffraction. The results show that the edge of the metal cone's base is the main scattering center of the structure. The intensity of the signal reflected from the lateral surface and vertex of the cone is much lower than the intensity of the field scattered by the edge of the cone's base.

CONCLUSION

Simulations have shown that, depending on the purpose of investigations, both time- and frequency-domain approaches appear to be efficient tools to analyze diffraction of nonstationary fields from various bodies.